\documentclass{aa}
\usepackage{txfonts}
\usepackage[dvips]{graphicx}
\usepackage{natbib}
\bibpunct{(}{)}{;}{a}{}{,}

\newcommand{\xte}{{\it RXTE} }
\newcommand{\pca}{{\it RXTE}/PCA }

\newcommand{\rg}{$R_{\rm g}$}
\newcommand{\hedge }{hard edge }

\begin{document}
\title{Vanishing hardness-flux correlation in Cygnus X-1: signs of the disc moving out}
\titlerunning{Vanishing hardness-flux correlation in Cyg X-1}
\authorrunning{M.~Axelsson et al.}
\author{Magnus Axelsson\inst{1} \and Linnea Hjalmarsdotter\inst{2,1} \and Luis Borgonovo\inst{3,1} \and Stefan Larsson\inst{1}}
\institute{Stockholm Observatory, AlbaNova, SE- 106 91 Stockholm, Sweden; magnusa@astro.su.se (MA) \and Observatory, P.O. Box 14, FIN-00014 University of Helsinki, Finland
\and Centre d'Etude Spatiale des Rayonnements, 31028 Toulouse, France}

\date{Received 9 May 2008 / Accepted 24 July 2008}

\abstract{}{We investigate observations of the X-ray binary Cygnus~X-1 with unusually high hardness and low flux. In particular, we study the characteristic frequencies seen in the PDS and the hardness-flux correlation within and between these observations.}{We analyse observations of Cyg~X-1 during periods when the source reaches its highest hardness levels ($\ga 1$ for the 9--20\,keV over 2--4\,keV \pca count ratios, corresponding to $\Gamma \la 1.6$).  Using the relativistic precession model to interpret the PDS, we estimate a value for the inner radius of the accretion disc. We also study the hardness-flux correlation.}{In the selected observations, the characteristic frequencies seen in the power spectrum are shifted to the lowest end of their frequency range. Within a single observation, the hardness-flux correlation is very weak, contrary to the negative correlation normally observed in the hard state. We suggest that this could be interpreted as the inner disc boundary being at large radii ($\ga$ 50 \rg), thereby requiring more time to adjust to a changing accretion rate than allowed by a single \textit{RXTE} observation, and compare our findings to estimates of the viscous time scale responsible for small scale variability in the system.}{}


\maketitle

\section{Introduction}
\label{intro}
\object{Cygnus~X-1} is one of the most studied X-ray sources, and is often quoted as the prototype black hole binary system. The source exhibits two main spectral states, commonly referred to as hard and
soft, with a brief intermediate state during transitions \citep[e.g., ][]{tl95,cui97,esi97,zg04}. Several models have been proposed to explain the observed states and transitions.
The two main components of such models are usually a geometrically
thin, optically thick accretion disk and a geometrically thick, optically thin hot inner flow or
corona. The models vary in the geometry and properties of mainly the
hot flow/corona \citep{haa93,bel99,cop99}. 

Transitions between the two states are believed to be a response to a change in accretion rate. In one widely accepted model \citep[see e.g., ][]{gie97,esi97,gie99,zdz02} these changes result in a reconfiguration of the accretion flow. In the soft state (high accretion rate) the disc extends almost in to the last stable orbit, whereas in the hard state (low accretion rate) it is truncated and replaced by a hot flow at inner radii. Throughout the hard state and into the transition, the hardness of the spectrum and changing temperature of the observed disc blackbody component is believed to track the variable radius of the inner disc. 

The energy spectrum of Cyg~X-1 in its hard state was described by \citet{gie97}. It is dominated by a component arising from thermal Comptonization in a plasma with electron temperature of $\sim 100$ keV and optical depth $\tau \sim 1$; only a weak blackbody component is visible, corresponding to an inner disc temperature of $\sim 200$ eV in a geometry with a truncated disc. An additional soft component, often modelled as an additional Comptonized component, is also required to accurately fit the spectrum.

Many sources also show a clear difference between the hard and soft states in the power density spectrum (PDS). In Cyg~\mbox{X-1}, the PDS in the soft state is generally well described by a single power law, with a cut-off at higher frequencies \citep{cui97}, while the PDS of the hard state shows a more complex structure, with several breaks or quasi-periodic oscillations \citep[QPOs; ][]{bh90,now99}.
The hard state PDS can be generally well-fit with broad Lorentzian components \citep{now00}. The number of components varies between studies, and depends on the frequency window and models used. \citet{pot03} conducted a large study using up to four Lorentzians in the 0.002--128\,Hz range. In \cite*{axe05} it was found that two Lorentzians are generally sufficient when fitting the PDS in the 0.01--25\,Hz range; by adding a cut-off power-law component to the model, the complete evolution of the PDS from hard to soft state could be modelled.

In this paper, we analyse observations of Cyg~X-1 in observations with unusually high hardness and low flux, the ``hard edge" of the hard state. In particular, we study the characteristic frequencies seen in the PDS and the hardness-flux correlation within and between these observations.

\begin{figure*}
\begin{centering}
\includegraphics[width=168mm]{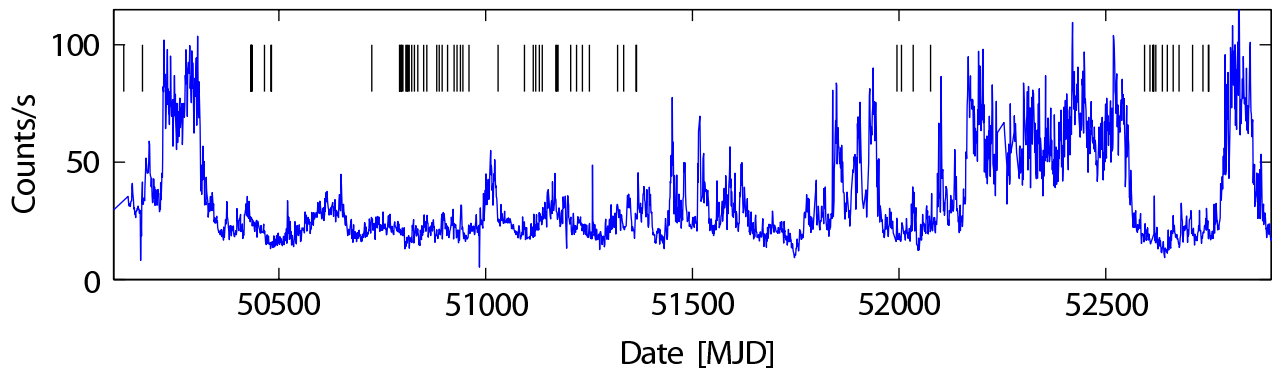}
\end{centering}
\caption{The one-day average ASM light curve for Cygnus~X-1. The hard and
soft states are clearly visible in the light curve, as well as a number of
large flares. The vertical lines in the figure mark the hard state episodes with unusually low flux analysed in this paper.}
\label{asm}
\end{figure*}

\section{Observations and data reduction}
Since its launch on December 30, 1995, the All-Sky Monitor \citep[ASM,][]{lev96} aboard
the {\it RXTE} satellite has provided nearly continual coverage of the X-ray sky in
the 2--12 keV range. A given source will be observed several times each
day, and the data is available either in dwell-by-dwell format, or one-day
averages. Figure \ref{asm} shows the one-day average lightcurve for Cygnus~X-1. The periods when the source enters the soft state are clearly visible, as well as a number of flaring episodes. There are also a number of occasions when the ASM flux drops below the normal hard state level, and these are
the observations studied in this paper. The data used are from pointed observations made with the PCA instrument onboard the \textit{RXTE} satellite \citep{jah96}. The times of the observations are indicated with vertical lines in Fig.~\ref{asm}. The typical length of an observation is $\sim0.5$--1 hour. 

To create the PDS, lightcurves were extracted with 10\,ms resolution in the 2--9\,keV band. The lightcurves were then separated in intervals of 8192 bins, and a PDS was computed for each interval. Only intervals without gaps were retained, and the resulting PDS were then averaged to form the final power spectrum. In calculating the hardness, PCA Standard2 data with 16\,s resolution were used. As the data stem from many different observations with varying setups, the PCUs in use vary between observations; all lightcurves have therefore been normalized to one PCU. The latest available bright source background model (dated August 6, 2006) was used. 

In the following analysis, we will concentrate on the observations with lowest flux, which also have the highest hardness levels, in agreement with the negative overall flux-hardness correlation in the hard state \citep[see for example][]{zdz02}. For easy reference, we will use the term ``hard edge'' to describe the episodes when the hardness ratio (HR) of Cyg~X-1 is roughly HR$\ga 1$ with the HR calculated as the count ratio between the 9--20\,keV and 2--4\,keV bands (corresponding to a spectral index $\Gamma \la 1.6$). 

\begin{figure}
\resizebox{\hsize}{!}{\includegraphics{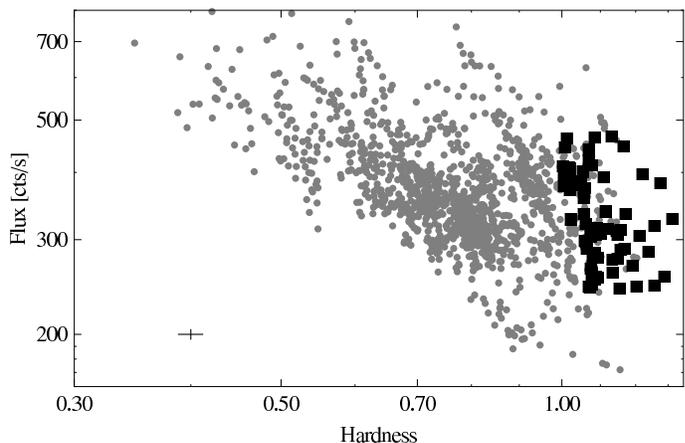}}
\caption{Hardness-flux correlation in Cyg~X-1 during the hard state. Black squares indicate the location of the observations where a third Lorentzian was required by \citet{axe05} to fit the PDS. The hardness is calculated for the 9--20\,keV over 2--4\,keV bands, and the flux is the sum of these two bands. Typical error bars are indicated in the lower left corner.}
\label{location}
\end{figure}

\section{Analysis and results}
\label{results}

\citet{gil99} found that the characteristic frequencies in the PDS of Cyg~X-1 correlated with the photon index. This was also seen in the systematic timing analysis of both \citet{pot03} and \citet{axe05}, where the peak frequencies of the Lorentzian components in Cyg~X-1 decreased with increasing hardness, indicating that the frequencies are shifting systematically. Indeed, a third QPO-type component was seen to enter the upper region of the frequency window for the hardest observations in the \citet{axe05} study, naturally explained as one of the higher frequency components seen in \citet{pot03}. Figure~\ref{location} shows a hardness-flux diagram of Cyg~X-1 in the hard state, with these observations represented by black squares. In all these observations, the two more common Lorentzian components are shifted to the lowest part of their frequency range.

\begin{figure}
\resizebox{\hsize}{!}{\includegraphics{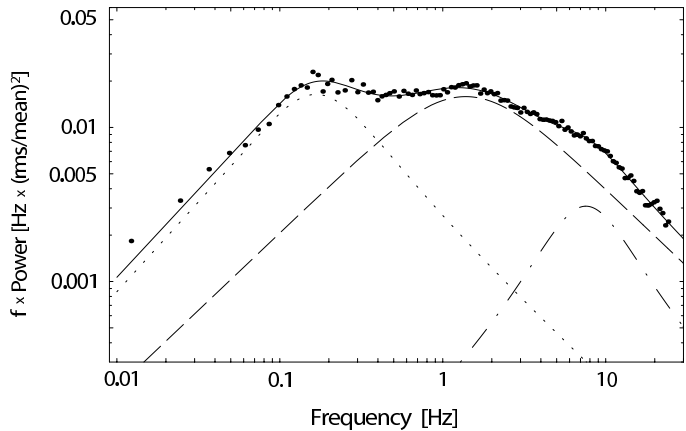}}
\caption{Power density spectrum of Cyg~X-1 in the hard edge. Three Lorentzian components are necessary to fit the data.}
\label{hardpowspec}
\end{figure}

In \citet{axe05}, the two hard state Lorentzian components were identified with relativistic precession frequencies \citep*{svm99} at the inner radius of a truncated disc. Shifts in frequency of the temporal components then correspond to changes in the truncation radius. Figure~\ref{location} clearly shows
that a shift to low frequencies, characterised by the third component appearing in the frequency window, are connected with high hardness -- all such observations occur in the hard edge of Cyg~X-1. 
An example \hedge PDS, taken from an observation on MJD 52\,748 (points) and fit using Lorentzian components (lines), is seen in Figure~\ref{hardpowspec}. 

\subsection{Estimating the inner disc radius}
\label{radius}

In the model used by \citet{gie97} \citep[\texttt{eqpair},][]{cop99} for the energy spectrum of \mbox{Cyg X-1}, a high hardness ratio reflects a high ratio of luminosity in the hot plasma electrons to that of the irradiating soft disc photons and a slightly lower disc temperature, suggesting a larger inner radius. The absolute values of disc temperature and inner radius from spectral modelling of data from the {\xte} are, however, not well constrained when the peak of the disc blackbody, as here, lies well below the lower boundary of the instrument ($\sim 3$ keV).

Many models of the fast variability in X-ray binaries tie the variations in flux to processes in the accretion disc. For example, disturbances can be introduced at large radii, and be modulated as they propagate inwards \citep{lyu97}. To give rise to a QPO, a given mode of variability must modulate the flux. In different models, the flux modulation is realised in different physical regions, and variability arising in the 
disc may therefore be observed at energies above those of the direct disc component.
  
Following \citet{axe05} we adopt the relativistic precession model. For this model to give rise to QPOs, a given precession frequency, and thereby a certain radius, must be preferred. 
In the disc truncation scenario, a natural candidate for such a radius is the boundary region between disc and hot inner flow. Variability transferred from the disc is thus observed as modulations in the Comptonized component. The frequency of these modulations is determined by the location of the boundary region, i.e. the inner radius of the disc. Adopting the frequency identification in \citet{axe05}, we can thereby calculate a value for the inner radius $R_{\rm in}$ of the accretion disc (in units of gravitational radii) using the observed frequencies:
\begin{equation}
R_{\rm in}\simeq 16.5 \left( \frac{M}{8\, M_\odot}\right)^{-\frac{2}{5}}\left( \frac{\nu_{\rm per}}{\rm 10\,Hz}\right)^{-\frac{2}{5}}\, , 
\end{equation}
\noindent where $M$ is the mass of the black hole and $\nu_{\rm per}$ the periastron precessional
frequency, here identified with the second Lorentzian component. This value is for all \hedge observations $\ga 50$ gravitational radii, indicating that the disc is truncated far from the innermost stable orbit. We note that this is in agreement with previous results by \citet{gil99} and \citet{ibr05}, who find correlations between reflection, spectral index and temporal characteristics: as the frequencies become lower, the spectrum hardens and reflection is reduced. 

The relativistic precession model identifies the two lower frequencies of the PDS with the nodal and periastron precessional frequencies, and a natural next step would be to try and identify the third component, or at least tie it to a physical process.

\begin{figure}
\resizebox{\hsize}{!}{\includegraphics{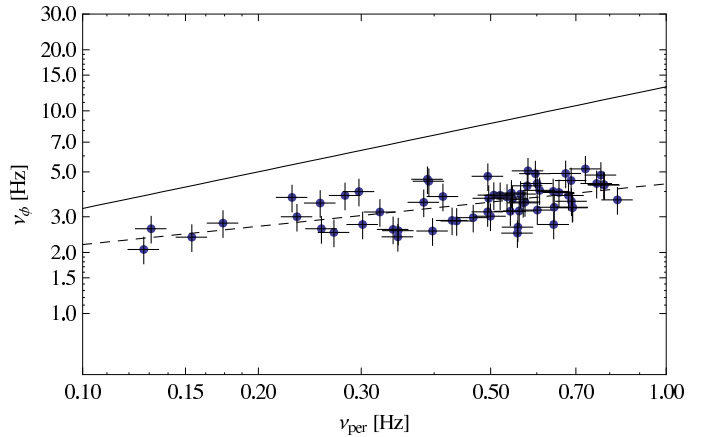}}
\caption{Correlation between frequencies of the second and third Lorentzians. The solid line indicates the orbital frequency ($\nu_{\phi}$) for a 10\,${\rm M}_{\odot}$ black hole as a function of periastron frequency ($\nu_{\rm per}$). The dashed line is the best-fit power-law to the data, yielding an index of $0.29\pm0.05$, whereas the orbital frequency is predicted to follow the relation $\nu_{\phi}\propto \nu_{\rm per}^{0.6}$.}
\label{orbtest}
\end{figure}

With the two lower frequencies assumed to arise from relativistic precession at the inner edge of the 
accretion disc, we may easily calculate the orbital frequency, $\nu_{\phi}$, as
\begin{equation}
\nu_{\phi}\simeq 33\;m^{-2/5}\nu_{\rm per}^{3/5}\; {\rm Hz},
\label{orbeq}
\end{equation}
\noindent where $m$ is the mass of the black hole in units of solar masses \citep{svm99}. Figure~\ref{orbtest} shows the measured frequency of the third Lorentzian component and predicted orbital frequency according to Eq.~\ref{orbeq} with a black hole mass of $10\;{\rm M}_{\odot}$.
The dashed line in the figure is the best-fit power-law to the data, giving 
an index of $0.29\pm0.05$. As can be seen, this is less steep than the 
prediction for the orbital frequency, $\nu_{\phi}\propto \nu_{\rm per}^{0.6}$.
The frequency of the third Lorentzian component thus appears to scale as $\nu_{\phi}^{0.5}$. In both the framework of relativistic precession and many other theoretical models for the QPOs, the observed frequencies are expected to vary with $\nu_{\phi}$; however, we have found no natural candidate for a frequency proportional to $\nu_{\phi}^{0.5}$ in any model. We stress that relativistic precession does not rule out additional components at higher frequencies; indeed, \citet{pot03} report a frequency component even higher than the ones discussed here. 

It is noteworthy that there has been no report of kHz variability in Cyg~X-1. If the identification here is correct, this absence is naturally explained, as such frequency ranges are far above what can be expected from the inner edge of the disc.

\subsection{Hardness-flux correlation}
An interesting feature of the hard edge observations is the behaviour of the hardness-flux correlation. As shown by several authors \citep[see for example][]{zha97,wen99,zdz02}, this correlation is negative in the hard state. This overall trend can be seen in Fig.~\ref{location}, although there is a large spread. 

\begin{figure}
\resizebox{\hsize}{!}{\includegraphics{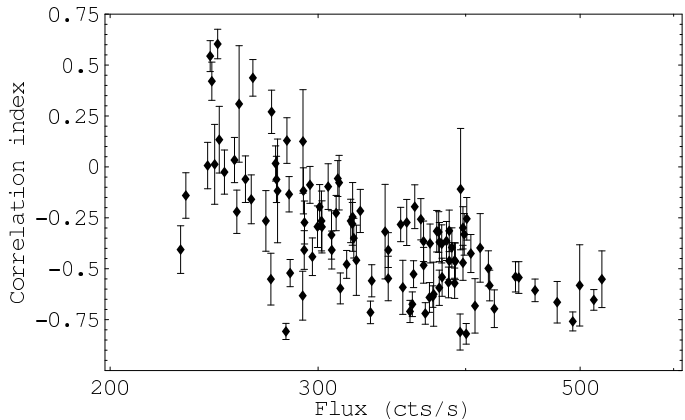}}
\caption{Spearman rank correlation index between hardness and flux as a function of flux. Each point corresponds to one observation of $\sim$30 minutes. The hardness is calculated for the 9--20\,keV over 2--4\,keV bands, and the flux is the sum of these two bands.}
\label{corrplot}
\end{figure}

In order to establish how the hardness-flux relation behaves in the \hedge cases, we calculate the Spearman rank correlation coefficient between hardness (9--20\,keV over 2--4\,keV) and flux for all our individual observations. 
The result is shown in Fig.~\ref{corrplot}.
\noindent
We find that as Cyg~X-1 enters the hard edge, the short term ($\sim$0.5--1 hr) negative correlation between hardness and flux (\textit{within} each observation) disappears. We note that the count rate remains above 100 counts/s in both channels, indicating that the effect is not artificially introduced. 

If we instead combine the observations to study the correlation on longer timescales, \textit{between} observations (from several hours to days), we find an overall hardness-flux correlation index of $-0.51\pm 0.09$. This shows that the correlation on longer timescales remains negative, even as it disappears within the individual observations. The uncertainty is estimated using the bootstrap method \citep[see, e.g.,][]{pre92}.

\begin{figure}
\resizebox{\hsize}{!}{\includegraphics{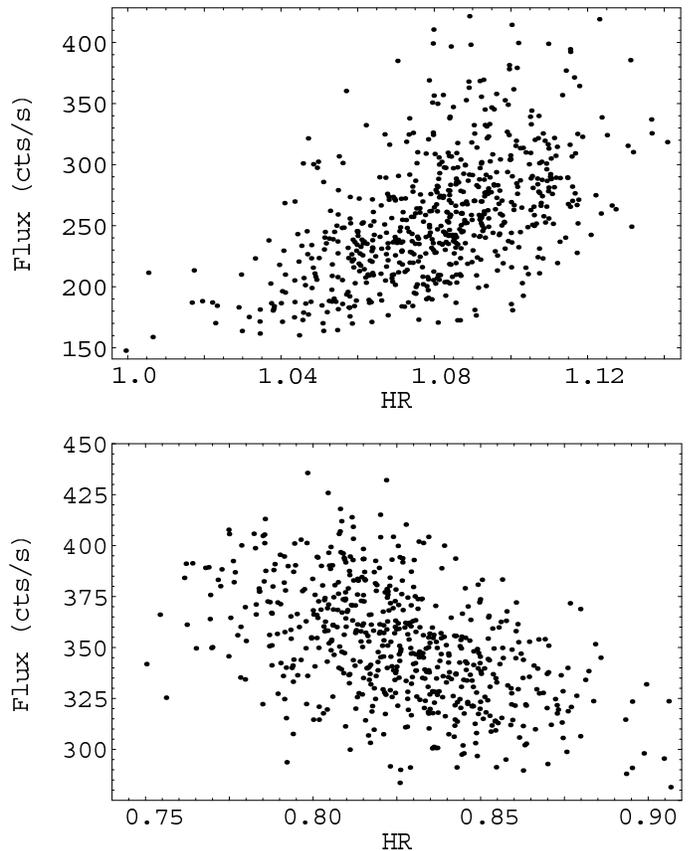}}
\caption{Correlation between hardness and flux in Cyg~X-1 in an observation from MJD 50\,481 (\textit{upper panel}) compared to a typical hard state observation from MJD 52\,693 (\textit{lower panel}). The correlation in the former observation is weakly positive, in contrast to the typical hard state relation. The hardness is calculated for the 9--20\,keV over 2--4\,keV bands, and the flux is the sum of these two bands.}
\label{hrflux}
\end{figure}

As can be seen in Fig.~\ref{corrplot}, there are a few observations where the hardness-flux correlation is actually positive. Looking closer at these observations, we find that they were all made within a 24-hour period of MJD 50\,481--50\,482. Figure~\ref{hrflux} shows
the hardness-flux correlation for one such observation, compared with that
more commonly seen in the hard state (from an observation on MJD 52\,693). 
\noindent
A Spearman rank test of the MJD 50\,481 observation yields the value $0.55\pm 0.04$, showing that the correlation is positive. In contrast, the same analysis for the normal hard state yields a value of $-0.43\pm 0.04$. 

To test how our choice of energy bands affects the calculated ratio, we extracted lightcurves in the bands matching the Standard2 quick-look lightcurves: 2--4\,keV, 2--9\,keV, 4--9\,keV, 9--20\,keV, and 20--40\, keV. This was done for both hard edge and typical hard state observations. All possible hardness ratios were then calculated. We find that using the 4--9\,keV over 2--4\,keV gives a weak correlation for both types of observations, probably since the bands are close and narrow. The typical hard state observation gave a negative correlation in all cases but one: when the highest ranges were used, 20--40\,keV over 9--20\,keV, the correlation was weakly positive. In the observations from MJD 50\,481--50\,482, a positive correlation was found for all combinations, but varying in strength. 

\section{Discussion}
\label{discuss}

\subsection{The radius of the inner disc}

Spectral studies are often performed on data which do not allow precise determination of the black body temperature. In addition, spectral modelling of several sources, including Cyg~X-1, require an additional soft component below 10 keV, making it difficult to determine the true temperature of the inner disc. It is therefore difficult to directly determine whether the inner radius of the accretion disc differs between spectral states. Although the disc truncation model has been successful in explaining the full range of states observed in X-ray binaries \citep{don03}, it is challenged by some observational results. Observation of broad iron lines \citep{mil02,min04,mil06b} and a strong black body spectral component \citep{mil06a,ryk07} both seem to indicate an untruncated disc in the hard state. However, these interpretations are strongly dependent on the detailed modelling of the continuum spectral shape, and currently there are no observations unambiguously conflicting with the truncated disc model \citep[see e.g.][]{don07}.

Timing studies offer an independent way of studying variations in the accretion flow between different states. Although models attempting to describe the PDS are less well established than spectral models, many connect observed frequencies with Keplerian motion in the accretion disc. Models such as relativistic precession allow a direct determination of the inner disc radius from characteristic frequencies in the PDS. Our results in Sect.~\ref{radius} indicate that the disc is truncated far ($\ga$ 50 \rg) from the innermost stable orbit in the hard edge observations. This supports the scenario where the spectrum becomes harder due to the inner disc receding, as a result of lower accretion rate. 

\subsection{The vanishing hardness-flux correlation - a case of mini-hysteresis?}
\citet{zdz02} found that the spectral variability of Cyg~X-1 within the hard state on long timescales could be described by a combination of two distinct variability patterns. In one of the patterns, the spectrum pivots at an energy around $\sim$50 keV, creating the observed flux-hardness anticorrelation. In the other, the spectrum varies in normalization without changes in the spectral shape. The two patterns were modelled by \citet{zdz02} using the Comptonization code \texttt{eqpair} \citep{cop99}. They found that the pivoting pattern could be explained by a variable input of soft seed photons to the Comptonizing flow. The varying strength of the spectrum with constant shape was interpreted as a variable bolometric luminosity caused by changes in the local accretion rate and optical depth of the hot flow, accompanied by a lower electron temperature to keep the Compton parameter $y\equiv 4\tau kT_e/m_ec^2$ constant (where $\tau$ is the optical depth and $T_e$ the electron temperature of the hot flow, and $m_e$ the electron mass).
\citet{mal06} found that the same two patterns could describe the variability on shorter timescales, from hours for the normalization pattern to days for the pivoting pattern, in an intermediate state of the source. 

Our data show that the anti-correlation between hardness and flux resulting from the pivoting behaviour still continues on long timescales into the highest hardness levels of Cyg~X-1, and suggest that our \hedge corresponds to an extremely low soft photon influx and a maximally truncated disc. The hardness-flux correlation within each observation shows that the pivoting behaviour is observable on timescales as short as 0.5--1 hour in the normal hard state. The vanishing hardness-flux correlation in the \hedge suggests that at this extreme, this variability pattern is not observed on such short timescales, even if it is still present on long timescales, between different observations. 

A possible interpretation of this behaviour is that when the disc is far away, $\ga$50 \rg, there is not enough time for the inner disc radius to adjust to a changing accretion rate within the time for a single \xte  observation. A change in the configuration of the accretion flow and the inner radius of the accretion disc is believed to take place on the viscous timescale \citep*{fkr},

\begin{equation}
t_{\rm visc} \sim \alpha^{-1} \left(\frac{H}{R}\right)^{-2} t_{\rm dyn} \, ,
\label{visceq}
\end{equation}

\noindent where $\alpha$ is the viscosity parameter, $H/R$ the ratio between disc scaleheight and radius, and $t_{\rm dyn}$ the dynamical timescale. Our frequency analysis suggests that the short term variability both in the normal hard state and in the \hedge is consistent with a variability of the inner radius of the order of a few \rg . If we assume that the varying influx of soft photons driving the pivoting behaviour is caused by the variability of the inner radius of the accretion disc we can calculate the expected timescale for this variability pattern. Assuming the dynamical timescale to be the orbital timescale in Fig.~\ref{orbtest} and inserting typical values \citep[$\alpha=0.1$, $H/R=0.01$;][]{mp07}, we can estimate the timescale of these variations of the inner disc radius to be of the order of 0.5--1 hour when the disc is truncated at $\sim$30 $R_{\rm g}$, consistent with the observed timescale for the pivoting pattern in the (normal) hard state.

From Eq.~(\ref{visceq}) it is evident that the viscous timescale scales directly with the dynamical timescale. Identifying the dynamical timescale with the orbital one gives: $t_{\rm visc} \propto t_{\rm \phi} \propto R^{3/2}$ \citep[for a thin disc with constant $\alpha$;][]{fkr}. A change in inner radius from $\sim$30 to $\ga$50 $R_{\rm g}$ in the \hedge would then result in a more than doubled timescale for the same fluctuations in the hard edge, compared to the normal hard state. This could then explain why we observe the hardness-flux correlation associated with the pivoting variability pattern within the time frame of a single \pca  observation in the normal hard state but not in the hard edge. Note that the existence of the anticorrelation between flux and hardness between different observations, separated by several hours to days, in the \hedge proves that the same mechanism is present -- an increase/decrease of the accretion rate is accompanied by a reconfiguration of the accretion disc with a change of the inner radius. However, the response at these large radii is too slow to be observed within the observation window of an \pca  observation. As a result, we may observe a change in flux within an observation, but the corresponding change in hardness only between different observations. 

The effect could be seen as a kind of ``mini-hysteresis" and compared to that observed in state transitions of certain X-ray transients. In e.g. GX~339-4, especially the hard-to-soft transition is observed at a wide range of luminosities \citep{zdz04}, seemingly in contradiction to a direct relationship between spectral state, luminosity and accretion rate. In some cases a luminosity as high as 30 per cent of the Eddington luminosity has been observed in the hard state before a transition takes place to the soft. It seems as if the local accretion rate increases much faster than the accretion flow is able to adapt and change structure from a hot flow to a cool disc. Since the source is a transient, at the beginning of an outburst the vicinity of the black hole is virtually devoid  of any material. In this situation the disc requires a longer time to build up, and the observed state is thus a function of both accretion rate and previous behaviour. In Cyg X-1, which is a persistent system where the accretion disc is always present, the transition between the hard and soft states takes place at the same luminosity and no such hysteresis effects are present. The effect seen in the hard edge, where we can observe changes in luminosity that are followed by a delayed change in the inner disc radius, can however be seen as a case of ``mini-hysteresis" in the sense of a delayed response of the disc to a change in accretion rate. 

Some of the observations presented here (MJD 50\,481--50\,482) show a positive hardness-flux correlation within one single observation. This behaviour is the exact opposite of the normal pivoting behaviour. Within the framework of \citet{zdz02}, such a variability pattern would be expected if the local accretion rate (and optical depth of the hot flow) increased while
the electron temperature remained unchanged. This could be the result of a momentary increase 
in accretion rate, on timescales shorter than that required for a general response in the accretion flow.

\section{Conclusions}
\label{conc}
We have studied observations of Cyg~X-1 made during hard-state episodes where  
the hardness ratio is at its highest. In particular, the PDS and the hardness-flux 
correlation have been compared to those of the hard state at usual hardness levels.
We find systematic differences in the \hedge cases: the PDS components are found at the lowest
end of their frequency range and the usually negative 
hardness-flux correlation weakens or even becomes positive within a single observation; however, the overall trend between different observations still follows the normal negative correlation, consistent with these harder energy spectra occurring predominantly during low flux episodes. 

We interpret these changes as indications of the inner disc being truncated at a large radius, $\ga50\;R_{\rm in}$. The third Lorentzian represents a frequency higher than the two generally observed, identified as the periastron and nodal frequencies in \citet{axe05}. This third frequency scales as $\nu_{\phi}^{0.5}$, and has no immediate physical interpretation. Our prediction for the orbital frequency also explains why no kHz QPOs are expected from Cyg~X-1. We relate the timescale for the observed short term pivoting  behaviour with the viscous timescale for variations of the inner radius of the accretion disc of a few \rg , and estimate that this may be more than twice as long in the \hedge than in the normal hard state, thus explaining the vanishing hardness-flux correlation in the hard edge. Our results support a framework where the hard state is explained by a truncated inner disc.

\begin{acknowledgements}
This research has made use of data obtained through the
High Energy Astrophysics Science Archive Research Center (HEASARC)
Online Service, provided by NASA/Goddard Space Flight Center. LH acknowledges support from the Academy of Finland, project nr. 1118854. We thank Juri Poutanen and Andrzej A. Zdziarski for helpful comments and discussions.
\end{acknowledgements}


\begin{thebibliography}{}

\bibitem[\protect\citeauthoryear{Axelsson, Borgonovo \& Larsson}{Axelsson et al.}{2005}]{axe05} Axelsson M., Borgonovo L., Larsson S., 2005, A\&A, 438, 999

\bibitem[\protect\citeauthoryear{Axelsson, Borgonovo \& Larsson}{Axelsson et al.}{2006}]{axe06} Axelsson M., Borgonovo L., Larsson S., 2006, A\&A, 452, 975 

\bibitem[\protect\citeauthoryear{Belloni \& Hasinger}{1990}]{bh90} Belloni T., Hasinger G., 1990, A\&A, 227, L33 

\bibitem[\protect\citeauthoryear{Beloborodov}{1999}]{bel99} Beloborodov, A.~M.\ 1999, 
ApJL, 510, L123 

\bibitem[\protect\citeauthoryear{Coppi}{1999}]{cop99} Coppi, P.~S.\ 1999, High Energy 
Processes in Accreting Black Holes, 161, 375

\bibitem[\protect\citeauthoryear{Cui et al.}{1997}]{cui97} Cui W., Zhang S.~N., Focke W., Swank J.~H., 1997, ApJ, 484, 383

\bibitem[\protect\citeauthoryear{Done \& Gierli{\'n}ski}{2003}]{don03} Done, C., \& Gierli{\'n}ski, M.\ 2003, MNRAS, 342, 1041 

\bibitem[\protect\citeauthoryear{Done, Gierli{\'n}ski \& Kubota}{Done et al.}{2007}]{don07} Done, C., Gierli{\'n}ski, M., \& Kubota, A.\ 2007, A\&ARv, 15, 1 

\bibitem[\protect\citeauthoryear{Esin, McClintock \& Narayan}{Esin et al.}{1997}]{esi97} Esin, A.~A., McClintock, J.~E., \& Narayan, R.\ 1997, ApJ, 489, 865 

\bibitem[\protect\citeauthoryear{Frank, King \& Raine}{Frank et al.}{2003}]{fkr} Frank J., King A., Raine D., 2003, Accretion Power in Astrophysics, Cambridge Universiy Press

\bibitem[\protect\citeauthoryear{Gierli{\'n}ski et al.}{1997}]{gie97} Gierlinski M., Zdziarski A.~A., Done C., Johnson W.~N., Ebisawa K., Ueda Y., Haardt F., Phlips B.~F., 1997, MNRAS, 288, 958 

\bibitem[\protect\citeauthoryear{Gierli{\'n}ski et al.}{1999}]{gie99} Gierli{\'n}ski M., Zdziarski A.~A., Poutanen J., Coppi P.~S., Ebisawa K., Johnson W.~N., 1999, MNRAS, 309, 496

\bibitem[\protect\citeauthoryear{Gilfanov, Churazov \& Revnivtsev}{1999}]{gil99} Gilfanov M., Churazov E., Revnivtsev M., 1999, A\&A, 352, 182 

\bibitem[\protect\citeauthoryear{Haardt \& Maraschi}{1993}]{haa93} Haardt, F., \& Maraschi, L.\ 1993, ApJ, 413, 507

\bibitem[\protect\citeauthoryear{Ibragimov et al.}{2005}]{ibr05} Ibragimov A., Poutanen J., Gilfanov M., Zdziarski A.~A., Shrader C.~R., 2005, MNRAS, 362, 1435 

\bibitem[\protect\citeauthoryear{Jahoda et al.}{1996}]{jah96} Jahoda, K., Swank, J.~H., Giles, A.~B., Stark, M.~J., Strohmayer, T., Zhang, W., \& Morgan, E.~H.\ 1996, Proc. SPIE, 2808, 59 

\bibitem[\protect\citeauthoryear{Kato}{2001}]{kat01} Kato S., 2001, PASJ, 53, 1

\bibitem[\protect\citeauthoryear{Levine et al.}{1996}]{lev96} Levine, A.~M., Bradt, H., Cui, W., Jernigan, J.~G., Morgan, E.~H., Remillard, R., Shirey, R.~E., \& Smith, D.~A.\ 1996, ApJL, 469, L33 
 
 \bibitem[\protect\citeauthoryear{Lyubarskii}{1997}]{lyu97} Lyubarskii, Y.~E.\ 1997, MNRAS, 292, 679 
 
\bibitem[\protect\citeauthoryear{Malzac et al.}{2006}]{mal06} Malzac J., et al., 2006, A\&A, 448, 1125 

\bibitem[\protect\citeauthoryear{Mayer \& Pringle}{2007}]{mp07} Mayer M., Pringle J.~E., 2007, MNRAS, 376, 435 

\bibitem[\protect\citeauthoryear{Miller et al.}{2002}]{mil02} Miller, J.~M., et al.\ 2002, ApJ, 578, 348

\bibitem[\protect\citeauthoryear{Miller et al.}{2006}]{mil06a} Miller, J.~M., Homan, J., \& Miniutti, G.\ 2006, ApJL, 652, L113

\bibitem[\protect\citeauthoryear{Miller et al.}{2006}]{mil06b} Miller, J.~M., Homan, 
J., Steeghs, D., Rupen, M., Hunstead, R.~W., Wijnands, R., Charles, P.~A., 
\& Fabian, A.~C.\ 2006, ApJ, 653, 525

\bibitem[\protect\citeauthoryear{Miniutti et al.}{2004}]{min04} Miniutti, G., Fabian, 
A.~C., \& Miller, J.~M.\ 2004, MNRAS, 351, 466 

\bibitem[\protect\citeauthoryear{Nowak et al.}{1999}]{now99} Nowak M.~A., Vaughan B.~A., Wilms J., Dove J.~B., Begelman M.~C., 1999, ApJ, 510, 874 


\bibitem[\protect\citeauthoryear{Nowak}{2000}]{now00} Nowak M.~A., 2000, MNRAS, 318, 361 

\bibitem[\protect\citeauthoryear{Pottschmidt et al.}{2003}]{pot03} Pottschmidt K., et al., 2003, A\&A, 407, 1039 

\bibitem[\protect\citeauthoryear{Press et al.}{1992}]{pre92} Press, W. H., Teukolsky, S. A., Vetterling, W. T., \& Flannery, B. P. 1992, Numerical Recipes in Fortran (2nd ed.; Cambridge: Cambridge Univ. Press )

\bibitem[\protect\citeauthoryear{Psaltis, Belloni \& van der Klis}{1999}]{psa99} Psaltis D., Belloni T., van der Klis M., 1999, ApJ, 520, 262 

\bibitem[\protect\citeauthoryear{Rykoff et al.}{2007}]{ryk07} Rykoff, E.~S., Miller, 
J.~M., Steeghs, D., \& Torres, M.~A.~P.\ 2007, ApJ, 666, 1129

\bibitem[\protect\citeauthoryear{Stella, Vietri \& Morsink}{Stella et al.}{1999}]{svm99} Stella L., Vietri M., Morsink S.~M., 1999, ApJ, 524, L63 

\bibitem[\protect\citeauthoryear{Tanaka \& Lewin}{1995}]{tl95} Tanaka Y., \& Lewin W. H. G., 1995, in X-Ray Binaries, ed. W. H. G. Lewin, J. van Paradijs, \& E. P. J. van den Heuvel (Cambridge: Cambridge Univ. Press), 126

\bibitem[\protect\citeauthoryear{van der Klis}{2004}]{vdk04} van der Klis, M., 2004, in Compact stellar X-ray sources, ed. W. H. G. Lewin \& M. van der Klis (Cambridge: Cambridge Univ. Press), 39 

\bibitem[\protect\citeauthoryear{Wagoner}{1999}]{wag99} Wagoner R.~W., 1999, PhR, 311, 259 

\bibitem[\protect\citeauthoryear{Wen et al.}{1999}]{wen99} Wen L., Cui W., Levine A.~M., Bradt H.~V., 1999, ApJ, 525, 968

\bibitem[\protect\citeauthoryear{Zdziarski et al.}{2002}]{zdz02} Zdziarski A.~A., Poutanen J., Paciesas W.~S., Wen L., 2002, ApJ, 578, 357 

\bibitem[\protect\citeauthoryear{Zdziarski \& Gierli{\'n}ski}{2004}]{zg04} Zdziarski A.~A., Gierli{\'n}ski M., 2004, PThPS, 155, 99 

\bibitem[\protect\citeauthoryear{Zdziarski et al.}{2004}]{zdz04} Zdziarski A.~A., Gierli{\'n}ski M., Miko{\l}ajewska J., Wardzi{\'n}ski G., Smith D.~M., Harmon, B. Alan, Kitamoto S., 2004, MNRAS, 351, 791 

\bibitem[\protect\citeauthoryear{Zhang et al.}{1997}]{zha97} Zhang S.~N., Cui W., Harmon B.~A., Paciesas W.~S., Remillard R.~E., van Paradijs J., 1997, ApJ, 477, L95

\end{thebibliography}
\end{document}